\documentclass[10pt,twocolumn,letterpaper]{article}

\usepackage{cvpr}
\usepackage{amsfonts}
\usepackage{multirow}
\usepackage[hidelinks,hypertexnames=false]{hyperref}
\usepackage{float}

\def\confName{CVPR}
\def\confYear{2026}
\def\paperID{0000}

\title{
Roadside LiDAR for Cooperative Safety Auditing at Urban Intersections: Toward Auditable V2X Infrastructure Intelligence
}

\author{
Bo Shang, Yiqiao Li\\
The City College of New York\\
{\tt\small \{bshang, yli4\}@ccny.cuny.edu}
}
\date{}

\begin{document}
\hypersetup{
  pdftitle={Roadside LiDAR for Cooperative Safety Auditing at Urban Intersections: Toward Auditable V2X Infrastructure Intelligence},
  pdfauthor={Bo Shang, Yiqiao Li}
}
\maketitle

\begin{abstract}
Urban intersections expose the limitations of single-vehicle perception under occlusion and partial observability. In this study, we present an auditable roadside LiDAR framework for infrastructure-assisted safety analysis at a signalized urban intersection in New York City, developed and evaluated using real-world data. The proposed framework integrates trajectory construction, iterative human-in-the-loop quality assurance (QA), and interpretable near-miss analytics to produce defensible safety evidence from infrastructure sensing. Using a human-labeled heavy vehicle--bicycle interaction as an anchor case, we show that direction-agnostic time-to-collision (TTC) drops below 1\,s, while longitudinal TTC remains above conservative braking thresholds, revealing a lateral-intrusion-dominated conflict mechanism. Beyond individual cases, continuous-window evaluation and multi-round QA analysis demonstrate that the framework systematically reduces failure modes such as track fragmentation, spurious TTC triggers, unstable geometry, and cross-lane false conflicts. These results position roadside LiDAR as a practical post-hoc auditing mechanism for cooperative perception systems, with broader statistical validation discussed. This work provides a pathway toward scalable, data-driven safety auditing of urban intersections, enabling transportation agencies to identify and mitigate high-risk interactions beyond crash-based analyses.
\end{abstract}

\section{Introduction}
Autonomous driving has advanced rapidly in 3D detection, tracking, and scene understanding, yet most deployed pipelines remain fundamentally single-vehicle-centric and therefore constrained by occlusion, limited viewpoints, and partial observability. These limitations are especially acute at urban intersections, where safety-critical interactions often unfold across crosswalks, turning lanes, and transient blind zones.

Vehicle-to-Everything (V2X) and infrastructure-assisted perception offer a complementary paradigm by extending observability beyond the ego vehicle. In this paper, we focus on a key subproblem within full-stack cooperative autonomy: how can roadside sensing provide \emph{auditable} and \emph{interpretable} safety evidence for reviewing safety-critical interactions at intersections? We study auditability and interpretability in the context of infrastructure-based evaluation, with a scope that focuses on evidentiary transparency and reviewability, rather than encompassing broader trustworthiness dimensions such as communication robustness, cybersecurity or regulatory compliance.

We investigate this question through a real-world pilot deployment of a roadside LiDAR--first safety-auditing pipeline at a site-specific urban intersection in New York City. The pipeline integrates 3D object detection, multi-object tracking, conservative trajectory refinement, a dynamics-aware post-tracking stabilization stage, and structured human-in-the-loop Quality Assurance (QA) to construct auditable trajectories for downstream surrogate safety analysis. The primary contribution lies in system design and validation methodology: how standard perception components are constrained, stabilized, reviewed, and interpreted to produce defensible safety evidence.

We use a human-labeled heavy vehicle--bicycle near-miss as an anchor case. Standard direction-agnostic Time-To-Collision (TTC) rapidly collapses below 1\,s, while longitudinal TTC relative to the heavy vehicle heading stays above conservative braking thresholds. This divergence suggests a lateral-intrusion-dominated interaction rather than insufficient longitudinal stopping capacity. Beyond this anchor case, we evaluate the tracking pipeline on continuous frame windows with stitched ground-truth trajectories and summarize repeated QA rounds that expose the dominant failure modes of infrastructure-side near-miss mining.

The paper makes four major contributions. First, we present an end-to-end roadside LiDAR pipeline for auditable safety analysis at a real-world urban intersection. Second, we establish an 8,000-frame manually annotated roadside LiDAR resource with frame-level cuboids plus later trajectory-level QA artifacts. Third, we demonstrate how structured review rounds transform raw detection and tracking outputs into a defensible and auditable near-miss analysis workflow. Fourth, we use longitudinal TTC, contrasted against direction-agnostic TTC, as an explanatory diagnostic for distinguishing braking-limited and lateral-intrusion-dominated interactions in roadside trajectory auditing.

\section{Related Work}
Cooperative perception and vehicle--infrastructure systems aim to reduce observability gaps through roadside sensing and shared scene context \cite{ji2024cvis_survey,huang2023v2xcp_survey,yazgan2024cp_datasets_survey}. Public datasets such as DAIR-V2X, V2X-Seq, TUMTraf-V2X, and UrbanIng-V2X have accelerated work on feature fusion, communication efficiency, and cooperative perception benchmarks \cite{yu2022dairv2x,yu2023v2xseq,zimmer2024tumtraf_v2x,urbaning2025urbaning_v2x}. In contrast, our work is single-site and narrower in scale, but it contributes a different artifact: a reviewed roadside LiDAR record built for safety auditing rather than only benchmark training and fusion evaluation.

Roadside LiDAR has also been used for intersection monitoring, traffic participant tracking, and occlusion-aware assistance \cite{roadside_lidar_review2026,lin2023roadside_tracking,mo2024enhanced_perception_intersections}. Meanwhile, surrogate safety measures such as TTC and Post-Encroachment Time (PET) remain standard tools for conflict analysis when crashes are too rare for direct estimation \cite{gettman2003ssm,hyden1987tct,johnsson2021relative_validation}. TTC itself has long been used in essentially longitudinal car-following settings, and later work generalized TTC to unconstrained two-dimensional motion \cite{jansson2005collision_avoidance,ward2015extending_ttc}. Our contribution is the use of longitudinal TTC, contrasted against direction-agnostic TTC, as an interpretive diagnostic within an auditable roadside trajectory pipeline. Despite recent advances, there remains a lack of an auditable pipeline that unifies roadside trajectory construction, structured human review, and near-miss interpretation into a transparent and reviewable artifact. This work seeks to fill that gap.
\begin{table*}[t]
\centering
\caption{Positioning of this work relative to representative infrastructure-side datasets.}
\label{tab:dataset_positioning_cr}
\footnotesize
\setlength{\tabcolsep}{3pt}
\begin{tabular}{p{0.13\textwidth}p{0.15\textwidth}p{0.13\textwidth}p{0.15\textwidth}p{0.13\textwidth}cp{0.18\textwidth}}
\toprule
Resource & Scale / coverage & Sensing setup & Annotation emphasis & Primary tasks & Audit/QA & Distinguishing scope \\
\midrule
DAIR-V2X \cite{yu2022dairv2x} & large real-world V2X benchmark & vehicle + infrastructure multi-view sensing & 3D object labels for cooperative perception & cooperative 3D detection & no & broader V2X benchmark scale; not organized around reviewed safety events \\
V2X-Seq \cite{yu2023v2xseq} & sequential real-world V2X scenes & vehicle + infrastructure sequential sensing & sequence labels supporting perception and prediction & cooperative perception, forecasting & no & temporal V2X benchmark; audit-oriented near-miss interpretation is not primary \\
TUMTraf-V2X \cite{zimmer2024tumtraf_v2x} & urban intersection traffic scenes & multi-sensor vehicle--infrastructure setup & scene-level labels for cooperative scene understanding & cooperative perception at intersections & no & intersection-focused benchmark with broader sensor stack, but not structured around safety auditing \\
UrbanIng-V2X \cite{urbaning2025urbaning_v2x} & broader multi-intersection benchmark & multi-site vehicle--infrastructure sensing & benchmark labels for cooperative perception & multi-intersection cooperative perception & no & larger geographic coverage and benchmark scope than this study \\
This work & single-site pilot roadside deployment & fixed roadside LiDAR-first sensing & 3D boxes, trajectories, reviewed near-miss cases, QA artifacts & detection, tracking, audited near-miss analysis & yes & narrower scale but unique emphasis on auditable trajectories, reviewed conflict interpretation, and hotspot-oriented safety analysis \\
\bottomrule
\end{tabular}
\normalsize
\end{table*}

\section{Methodology}
\begin{figure*}[t]
\centering
\includegraphics[width=0.7\textwidth]{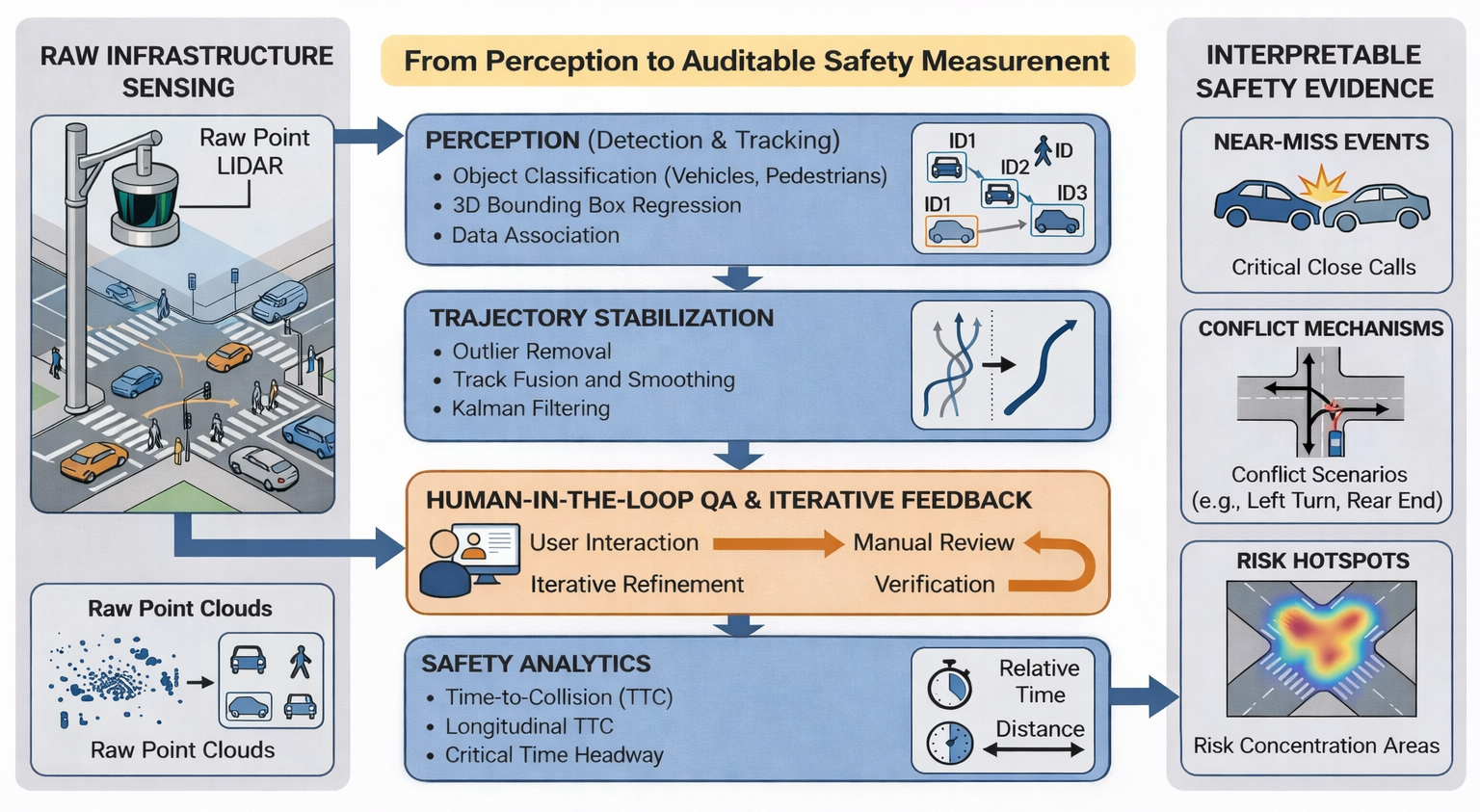}
\caption{Conceptual overview of the roadside LiDAR safety-auditing architecture. Raw point clouds are converted into detections, trajectories, and stabilized motion estimates, then reviewed through a human-in-the-loop QA layer and interpreted with surrogate-safety analytics to produce auditable safety evidence such as reviewed near-miss cases, conflict mechanisms, and risk hotspots.}
\label{fig:system_overview_cr}
\end{figure*}

Figure~\ref{fig:system_overview_cr} summarizes the methodology as a perception-to-audit workflow. The perception stack covers Stage 1 detections, Stage 2 identity-consistent tracks, and Stage 3 refined trajectories; human-in-the-loop review then acts as a supervisory validation layer before and during Stage 4 near-miss analytics, which produces the final auditable safety evidence.

\subsection{Stage 1: 3D detection.}
Each frame is processed using a fine-tuned CenterPoint detector \cite{yin2021centerpoint} trained on our 8,000 manually annotated roadside LiDAR dataset. For each detected object, the model outputs a class label $c$, confidence score $s$, and 3D bounding box parameters $(x, y, z)$ (center coordinates), $(d_x, d_y, d_z)$ (box dimensions), and $\psi$ (heading angle).

\subsection{Stage 2: Multi-object tracking for Data Association.}
To establish temporal correspondence, detections are associated across frames using a SORT-style tracking framework \cite{bewley2016sort} that integrates Kalman filtering with bird’s-eye-view (BEV) gating. This step produces stable and persistent track identities, forming the basis for reliable trajectory-level analysis.

\subsection{Stage 3: Refinement and stabilization.}
While multi-object tracking provides temporally consistent identities, the resulting trajectories often exhibit frame-level noise, orientation jitter, and occasional misalignment due to detection uncertainty and association errors. This noise can propagate into downstream safety metrics (e.g., TTC) and compromise interpretability. Therefore, a refinement stage is necessary to improve temporal stability while preserving auditability.

{\it Branch-level refinement policy.} To keep the audited comparisons controlled, we maintain three post-tracking refinement branches under identical upstream detections and audited frame windows. \textbf{B0} retains the raw tracked trajectories as a baseline. \textbf{B1} introduces a selective correction mechanism that targets potentially unreliable track segments. Suspicious tracklets are identified based on three criteria: (i) \emph{yaw-step discontinuities}, defined as abrupt frame-to-frame changes in the estimated heading angle $\psi_t$ that exceed a predefined threshold and are inconsistent with physically plausible motion; (ii) \emph{registration disagreement}, measured as the mismatch between the predicted state from the tracker (e.g., Kalman filter extrapolation) and the observed detection, typically quantified by BEV spatial residuals or IoU-based inconsistencies; and (iii) a \emph{thresholded suspicion score}, which aggregates normalized indicators (e.g., yaw variation, position error, and detection confidence drops) into a composite reliability score used to flag unstable track segments. For tracklets flagged as suspicious, \textbf{B1} applies \emph{capped frame-level corrections}, where adjustments to position and orientation are bounded to prevent over-smoothing or distortion of genuine motion patterns. These corrections are only applied when \emph{registration quality is deemed sufficient}, meaning that the detection-to-track alignment satisfies minimum criteria (e.g., IoU above a threshold or residual below a tolerance), ensuring that corrections are anchored to reliable observations rather than propagating tracking errors. \textbf{B2} applies stronger frame-wise temporal smoothing across the entire trajectory, enforcing higher temporal consistency but at the potential cost of attenuating sharp but valid motion changes. 

We use \textbf{B1} as the practical default because it preserves audited pair-level summaries while modestly reducing orientation jitter. Longitudinal TTC is then applied to the selected audited pairs as an interpretive diagnostic rather than a standalone binary event threshold.

{\it Dynamics-aware stabilization extension.} Beyond the B0/B1/B2 branch comparison, we also evaluate a separate dynamics-aware stabilization module. This extension is intended to suppress high-frequency noise while enforcing physically plausible motion, but because it can materially change the recovered anchor-case geometry, we report it separately rather than fold it into the default audited branch. Let $\mathbf{p}_t = (x_t, y_t)^\top$ denote the observed box center (BEV center), $\psi_t$ the heading, and $\mathbf{d}_t = (d^{\mathrm{aligned}}_{x,t}, d^{\mathrm{aligned}}_{y,t}, d_{z,t})^\top$ the aligned dimensions. Let $\mathcal{W}_t$ denote a centered temporal window.

{\it Position smoothing and stabilization.}
A local estimator is first constructed as a weighted average:
\begin{equation}
\bar{\mathbf{p}}_t =
\frac{\sum_{\tau \in \mathcal{W}_t} w_{t,\tau}\,\mathbf{p}_\tau}
{\sum_{\tau \in \mathcal{W}_t} w_{t,\tau}}, 
\qquad w_{t,\tau} \ge 0,
\end{equation}
where $w_{t,\tau}$ denotes the temporal smoothing weight. In the current implementation,
$w_{t,\tau}=1$ for frames $\tau \in \mathcal{W}_t$ and $w_{t,\tau}=0$ otherwise, yielding a centered
moving average with a default window size of 9 frames. The smoothed position is then passed through
a separate constant-velocity prior, so temporal averaging and motion regularization are applied in
two successive steps.

Let $\tilde{\mathbf{p}}_t$ denote the stabilized position and define the finite-difference velocity
\begin{equation}
\mathbf{v}_{t-1} = \frac{\tilde{\mathbf{p}}_{t-1} - \tilde{\mathbf{p}}_{t-2}}{\Delta t}.
\end{equation}
A constant-velocity prediction is given by
\begin{equation}
\hat{\mathbf{p}}_t = \tilde{\mathbf{p}}_{t-1} + \mathbf{v}_{t-1}\Delta t.
\end{equation}
The stabilized estimate is obtained via convex combination:
\begin{equation}
\tilde{\mathbf{p}}_t = (1-\alpha)\,\bar{\mathbf{p}}_t + \alpha\,\hat{\mathbf{p}}_t,
\quad \alpha \in [0,1].
\end{equation}

{\it Heading stabilization.}
To account for angular periodicity, heading is smoothed using circular averaging:
\begin{equation}
\bar{\psi}_t =
\operatorname{atan2}\!\left(
\sum_{\tau \in \mathcal{W}_t} w_{t,\tau}\sin \psi_\tau,\,
\sum_{\tau \in \mathcal{W}_t} w_{t,\tau}\cos \psi_\tau
\right).
\end{equation}
Equation (5) computes a weighted circular mean of yaw angles over a temporal window by averaging their sine and cosine components and recovering the orientation via the $atan2$ function. This formulation avoids discontinuities due to angular wrap-around and yields a stable and physically consistent estimate of heading.
Let $\psi^{\mathrm{motion}}_t$ denote the motion-induced heading (e.g., path tangent). Define the wrapped angular difference
\[
\Delta\psi(a,b) := \operatorname{wrap}(a-b) \in (-\pi,\pi].
\]
A blended heading is first computed as
\begin{equation}
\psi_t^{\mathrm{blend}} =
\bar{\psi}_t + \beta\,\Delta\psi(\psi^{\mathrm{motion}}_t,\bar{\psi}_t),
\quad \beta \in [0,1].
\end{equation}
In the current implementation, $\beta$ is adaptive rather than fixed. Let $s_t$ denote speed,
$s_{\min}$ the minimum speed at which motion heading becomes reliable, and let $r_t$ denote the
resulting heading-reliability coefficient:
\begin{equation}
\begin{aligned}
r_t &= \min\!\left(\frac{s_t-s_{\min}}{s_{\min}+0.75},\,1\right),\\
\alpha_t &= \alpha_{\mathrm{low}}+(\alpha_{\mathrm{high}}-\alpha_{\mathrm{low}})\,r_t.
\end{aligned}
\end{equation}
Here $\alpha_t$ denotes the implementation-side raw-yaw blend factor, while $r_t$ measures
heading reliability from motion evidence.
with defaults $\alpha_{\mathrm{low}}=0.35$, $\alpha_{\mathrm{high}}=0.08$, and
$s_{\min}=0.75\,\mathrm{m/s}$. Because the implementation parameterizes the blend from the
motion-heading side, the equivalent coefficient in Eq.~(6) is
\begin{equation}
\begin{aligned}
\beta_t &=
\begin{cases}
1-\alpha_t, & \text{if } |\Delta\psi(\bar{\psi}_t,\psi_t^{\mathrm{motion}})| \le \epsilon_\psi,\\
1, & \text{otherwise,}
\end{cases}\\
\epsilon_\psi &= 20^\circ.
\end{aligned}
\end{equation}
Thus $\beta_t$ increases with motion reliability: at low speed the blend remains closer to the raw
smoothed heading, while at higher speed it shifts toward the motion-induced heading.

To enforce a bounded turn rate, the update is constrained by projection onto the admissible increment set:
\begin{equation}
\tilde{\psi}_t =
\tilde{\psi}_{t-1}
+
\Pi_{[-\omega_{\max}\Delta t,\ \omega_{\max}\Delta t]}
\bigl(\Delta\psi(\psi_t^{\mathrm{blend}}, \tilde{\psi}_{t-1})\bigr),
\end{equation}
where $\omega_{\max}$ is the maximum allowed angular velocity, so the proposed heading increment is
clipped to the interval $[-\omega_{\max}\Delta t,\ \omega_{\max}\Delta t]$: the lower bound
$-\omega_{\max}\Delta t$ is the largest allowed clockwise step and the upper bound
$+\omega_{\max}\Delta t$ is the largest allowed counterclockwise step. Here
$\Pi_{[a,b]}(x)=\min(\max(x,a),b)$ denotes scalar clipping onto $[a,b]$. In the current
implementation this is realized as a per-frame angular-step limit, with default
$\omega_{\max}\Delta t = 2.5^\circ$ for consecutive frames and proportional scaling for larger frame gaps.

Optionally, $\tilde{\psi}_t$ is back-propagated from the first reliable-motion frame to stabilize low-speed initialization.

{\it Dimension stabilization.}
Object dimensions are stabilized componentwise via robust estimation:
\begin{equation}
\tilde d_i = \arg\min_{d} \sum_t \rho(d_{i,t} - d), 
\quad i \in \{x,y,z\},
\end{equation}
and $\tilde{\mathbf d} = (\tilde d_x,\tilde d_y,\tilde d_z)^\top$, where $\rho(\cdot)$ is a convex robust loss (e.g., $\ell_1$). $d_{i,t}$ is the observed size at time $t$. $d$ is candidate "true" dimension.

\subsection{Cross-Cutting Audit Layer: Human QA.}
Automated review queues are exported as short tracklet windows with aligned point-cloud context, box overlays, and candidate-event metadata so that reviewers inspect the same evidence used by downstream mining. Each queued item is checked for identity continuity, pose plausibility, split/merge failures, and whether the apparent event reflects a true conflict, a TTC misuse case, or a geometry / lane-assignment artifact. Reviewer outcomes are stored as structured correction records rather than free-form notes, including keep / reject decisions, dominant failure tags, and any required track-level corrections. These records are then fed back into the pipeline in two ways: they correct the audited track outputs for reported examples, and they update the near-miss screening logic by exposing recurring false-positive modes such as fragmented tracks, invalid closing-motion assumptions, and unstable box geometry.

\subsection{Stage 4: Near-Miss Analytics.}
Candidate pairs are first screened by direction-agnostic TTC and minimum separation, then interpreted with longitudinal TTC to distinguish longitudinal braking-limited risk from lateral-intrusion-dominated interactions. We use TTC as a standard surrogate safety measure from the traffic-conflict literature \cite{gettman2003ssm,hyden1987tct,johnsson2021relative_validation}, while minimum separation is treated here as a practical implementation-specific BEV clearance proxy rather than a canonical published metric. For a candidate pair indexed by $i\in\{1,2\}$, let $\Delta \mathbf{p}=\mathbf{p}_2-\mathbf{p}_1$, $\Delta \mathbf{v}=\mathbf{v}_2-\mathbf{v}_1$, and $r_i=\tfrac{1}{2}\sqrt{d_{x,i}^2+d_{y,i}^2}+\beta$, where $d_{x,i},d_{y,i}$ are the BEV box dimensions of participant $i$, $\beta$ is a fixed radius buffer (0.3\,m in the current implementation), and $t$ indexes frames within the interaction window. We use
\begin{align}
s_t &= \|\Delta \mathbf{p}(t)\| - \bigl(r_1(t)+r_2(t)\bigr), \\
\mathrm{TTC}(t) &=
\begin{cases}
\dfrac{-\Delta \mathbf{p}(t)\cdot \Delta \mathbf{v}(t)}{\|\Delta \mathbf{v}(t)\|^2}, & \Delta \mathbf{p}(t)\cdot \Delta \mathbf{v}(t) < 0,\\
+\infty, & \text{otherwise},
\end{cases}
\end{align}
and summarize each candidate pair by $\min_t s_t$ and $\min_t \mathrm{TTC}(t)$ over the interaction window. Negative minimum-separation values therefore indicate overlap of the size-adjusted BEV proxies, not literal physical penetration.

\section{Experimental Setup}
\paragraph{Study site and sensing.}
Data were collected at the signalized intersection of Convent Ave \& W 141st St near the City College of New York using a fixed roadside Ouster OS-1-128 LiDAR operating at 10\,Hz. The deployment is a fixed elevated oblique-view setup covering the intersection center and crosswalk approaches. All object positions, trajectories, and near-miss measurements are reported in one common BEV coordinate system aligned to this intersection and held fixed throughout the study.

\begin{figure}[t]
\centering
\includegraphics[width=0.9\linewidth]{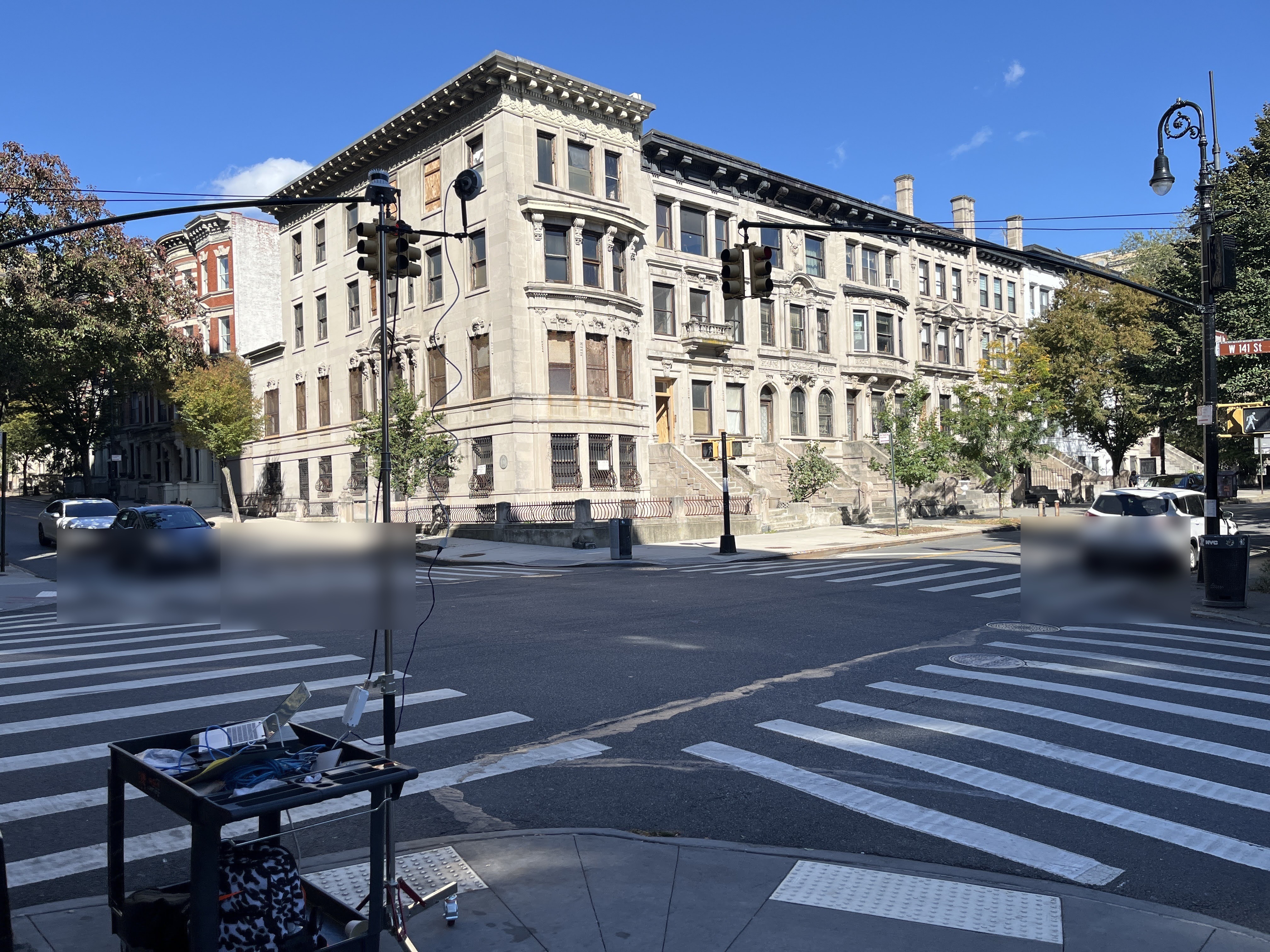}
\caption{Study intersection and roadside sensing setup used for the pilot deployment.}
\label{fig:data_collection_site_cr}
\end{figure}

\paragraph{Annotation resource.}
The 8,000 study set consists of frame-level 3D cuboid annotations for four coarse classes: car, truck, pedestrian, and bicycle. Source annotations were created in MATLAB LiDAR Labeler at the native 10\,Hz frame rate, then converted into the formats used for detector training and audit evaluation. Quality control has two stages: human-reviewed ground-truth frame-label review for detector evaluation and later trajectory-level re-review for audited windows. This differs from benchmark-only infrastructure datasets by combining frame labels with later trajectory-QA artifacts and reviewed near-miss analysis.

\paragraph{Evaluation protocol.}
We report three forms of evidence: detector performance against human-reviewed ground truth, branch-level tracking and refinement comparisons on audited and continuous frame windows with stitched ground-truth trajectories, and human cross-checks for reviewed near-miss candidates. The present paper is a single-site pilot, so these results should be interpreted as feasibility and auditability evidence rather than a comprehensive generalization study.

This deployment choice is intentional: by fixing the sensor, calibration frame, and site geometry, the study establishes a controlled audit reference rather than a general-purpose multi-site benchmark. The trade-off is that synchronization error, calibration maintenance across many intersections, and bandwidth-constrained V2X delivery are left to future systems work.

\begin{figure}[t]
\centering
% Source frame for reproducibility: audited interaction window, frame 6532.
\includegraphics[width=\linewidth]{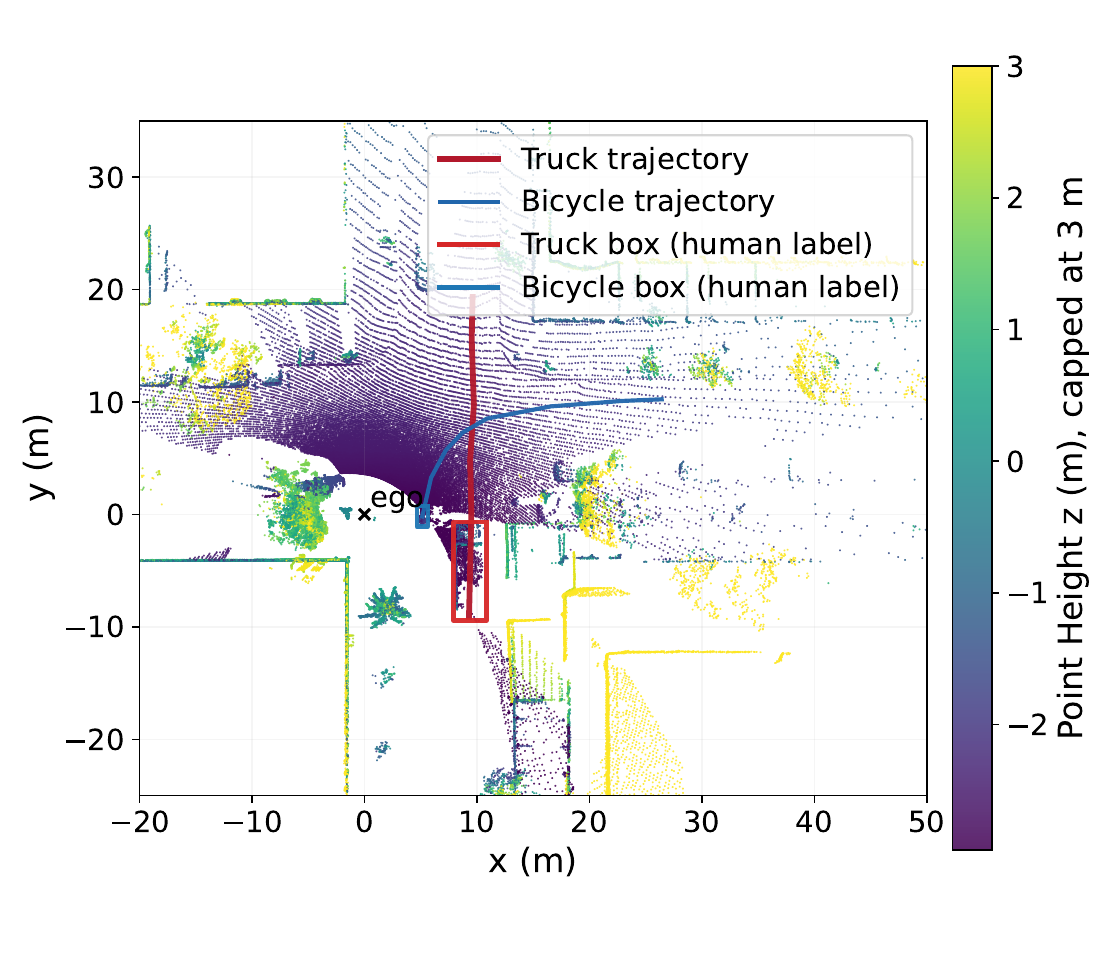}
\caption{Representative roadside LiDAR point cloud from the audited interaction window, shown in bird's-eye view with human-labeled truck and bicycle cuboids.}
\label{fig:pointcloud_bev_example_cr}
\end{figure}

\begin{figure}[t]
\centering
\includegraphics[width=\linewidth]{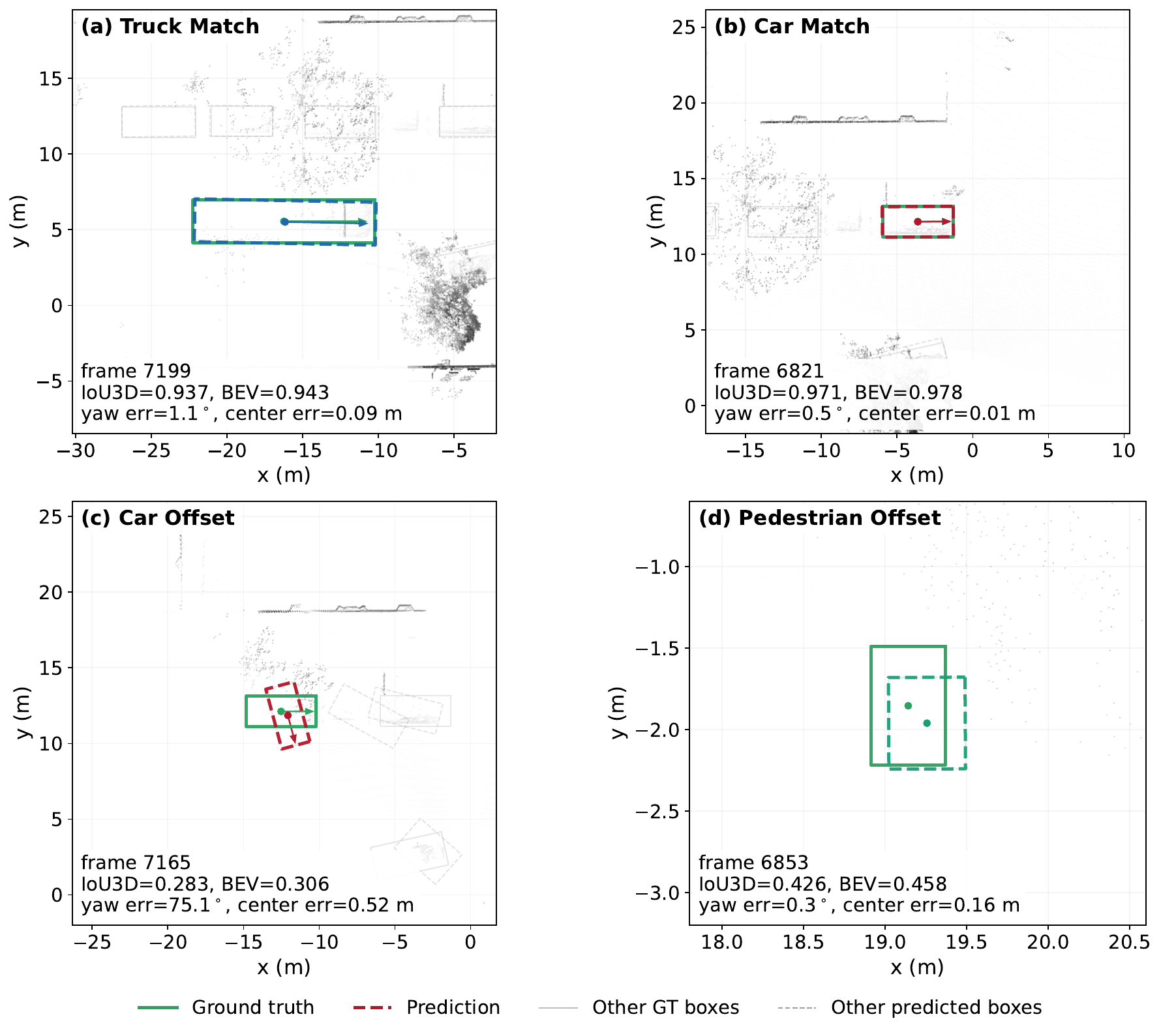}
\caption{Qualitative detection examples from the CenterPoint model trained on the 8,000-frame dataset, evaluated on the consecutive validation split: (a) a high-overlap truck match, (b) a high-overlap car match, (c) a car case with noticeable localization offset, and (d) a single-pedestrian case with moderate localization offset. Car and truck detections are generally strong, while small-object localization remains more fragile.}
\label{fig:detection_demo_cr}
\end{figure}

\section{Results}
\subsection{Detection, Tracking, and QA}
\begin{table}[t]
\centering
\caption{Compact quantitative summary of the main pipeline and audit-layer components.}
\label{tab:main_quant_summary_cr}
\footnotesize
\setlength{\tabcolsep}{3pt}
\begin{tabular}{p{0.34\linewidth}p{0.62\linewidth}}
\toprule
Component & Result \\
\midrule
Detector (human-reviewed GT holdout) & Vehicle AP@0.70 (R40) 76.38\%; native AP: car 89.75\%, truck 54.55\%, pedestrian 60.10\%; 38.31 fps \\
B0 baseline refinement & F1 0.9825; yaw$_{p95}$ 15.19$^\circ$ \\
B1 registration-guided refinement & F1 0.9825; yaw$_{p95}$ 14.04$^\circ$ \\
B2 temporal smoothing & F1 0.9780; yaw$_{p95}$ 3.92$^\circ$ \\
Dynamics-aware stabilization & Heading-motion error 10.45$^\circ \rightarrow$ 2.36$^\circ$; heading-step 0.0251 $\rightarrow$ 0.0040 rad \\
\bottomrule
\end{tabular}
\normalsize
\end{table}

\begin{table}[t]
\centering
\caption{Tracking evaluation on continuous frame windows against ground-truth trajectories stitched across consecutive frames. A predicted box is counted as a match when its BEV center lies within 1.5\,m of the corresponding ground-truth box center.}
\label{tab:stitched_window_eval_cr}
\footnotesize
\setlength{\tabcolsep}{4pt}
\begin{tabular}{lcccc}
\toprule
Window & Precision & Recall & F1 & yaw$_{p95}$ \\
\midrule
First 2000 frames & 0.6947 & 0.9858 & 0.8150 & 3.88$^\circ$ \\
First 4000 frames & 0.5734 & 0.8677 & 0.6905 & 6.50$^\circ$ \\
Sequential validation slice & 0.6036 & 0.6726 & 0.6362 & 3.47$^\circ$ \\
\bottomrule
\end{tabular}
\normalsize
\end{table}

\begin{table}[t]
\centering
\caption{Human cross-check summary for near-miss validation.}
\label{tab:near_miss_crosscheck_cr}
\footnotesize
\setlength{\tabcolsep}{4pt}
\begin{tabular}{p{0.52\linewidth}p{0.18\linewidth}p{0.20\linewidth}}
\toprule
Cross-check item & Result & Evidence \\
\midrule
Completed dashboard review decisions & 26 & rounds 000--003 \\
Confirmed true near-miss decisions & 1 & round 002 \\
Provisional / borderline positives & 1 & round 001 \\
Deferred decisions pending better geometry & 1 & round 002 \\
Anchor-case class-pair agreement & yes & truck--bicycle \\
\bottomrule
\end{tabular}
\normalsize
\end{table}

\begin{table}[t]
\centering
\caption{Iterative trajectory-QA validation rounds for refining the near-miss review pipeline.}
\label{tab:qa_rounds_cr}
\footnotesize
\setlength{\tabcolsep}{2pt}
\renewcommand{\arraystretch}{1.05}
\begin{tabular}{p{0.10\linewidth}p{0.12\linewidth}p{0.22\linewidth}p{0.46\linewidth}}
\toprule
Round & Cases & Issue & Main change or finding \\
\midrule
000 & 10 & tracking breaks & No true near-miss; track fragmentation filtered before metric tuning. \\
001 & 10 & TTC misuse / weak distance & Motion filtering and TTC gating reduced over-triggering in near-static or yielding cases. \\
002 & 3 & TTC / clearance ambiguity & Anti-repeat sampling removed replay pairs; one near-miss remained but clearance was unreliable. \\
003 & 3 & box geometry / lane conflict & Stationary-aware TTC and BEV clearance reduced false positives; geometry still unstable. \\
\bottomrule
\end{tabular}
\normalsize
\end{table}

Table~\ref{tab:main_quant_summary_cr} shows that the paper's strongest quantitative support is not detector novelty but the combination of reviewed trajectory construction and controlled refinement. B1 preserves the audited F1 of the baseline while modestly reducing yaw jitter, whereas B2 achieves stronger smoothing at a small cost in F1. The dynamics-aware branch further improves geometric consistency, but because it materially changes the anchor near-miss geometry, we report it as a separate engineering extension rather than substitute it into the core audited comparison.

Figure~\ref{fig:detection_demo_cr} is consistent with a local-domain deployment story rather than cross-site generalization: the detector localizes cars and trucks well in the study geometry, but pedestrian-scale and weak-overlap cases remain more brittle. Table~\ref{tab:stitched_window_eval_cr} broadens the evidence beyond the anchor interaction. Performance is strongest on the shorter first-2000 slice and declines on longer windows, mainly through accumulated false positives and missed long-tail cases rather than catastrophic geometric instability. The iterative QA process in Table~\ref{tab:qa_rounds_cr} is equally important: dominant failure modes evolved from track fragmentation to TTC misuse, then to geometry instability and different-lane false conflicts. The public QA dashboard records 26 completed review decisions across rounds 000--003.\footnote{\href{https://trajectory-qa.aimobilitylab.xyz/}{trajectory-qa.aimobilitylab.xyz/}}

These results also clarify what is and is not being claimed. We do not present an apples-to-apples leaderboard comparison against recent cooperative fusion methods because those works use different sensing stacks, synchronization assumptions, communication protocols, and target tasks. Instead, the present evaluation focuses on what is directly valid under this roadside audit setup: detector performance against human-reviewed ground truth, controlled branch-level refinement trade-offs, broader continuous-window tracking behavior with stitched ground-truth trajectories, and reviewed model-versus-human near-miss timing agreement.

\subsection{Near-Miss Interpretation}
\begin{table}[t]
\centering
\caption{Anchor near-miss comparison for the same truck--bicycle interaction. Minimum separation is a signed size-adjusted clearance; negative values in $\Delta$ indicate that the model predicts a tighter or more overlapping interaction than the human reference.}
\label{tab:anchor_compare_cr}
\footnotesize
\setlength{\tabcolsep}{4pt}
\begin{tabular}{lccc}
\toprule
Metric & Human & Model & $\Delta$ \\
\midrule
Min TTC (s) & 0.62 & 0.55 & -0.07 \\
Min separation (m) & 0.81 & 0.20 & -0.61 \\
Min center distance (m) & 6.47 & 5.61 & -0.86 \\
Min TTC timing & reference & +2 frames & +0.2 s \\
Min sep. timing & reference & +2 frames & +0.2 s \\
\bottomrule
\end{tabular}
\normalsize
\end{table}

\begin{table}[t]
\centering
\caption{Multi-case audited near-miss summary from the same run under the final conservative trajectory pipeline. `Min sep.' denotes signed size-adjusted clearance, so negative values indicate estimated overlap under the box proxy.}
\label{tab:multi_case_audit_summary_cr}
\footnotesize
\setlength{\tabcolsep}{3pt}
\begin{tabular}{lccc}
\toprule
Case & Pair & Min TTC (s) & Min sep. (m) \\
\midrule
01 & truck-bicycle & 0.552 & 0.201 \\
02 & truck-truck & 0.131 & -4.017 \\
03 & car-bicycle & 0.455 & 5.870 \\
04 & car-car & 2.563 & 12.563 \\
05 & truck-car & 1.249 & 5.009 \\
\bottomrule
\end{tabular}
\normalsize
\end{table}

For the anchor truck--bicycle interaction, the model recovers the same critical timing within two frames of the human reference (Table~\ref{tab:anchor_compare_cr}) but estimates a tighter minimum gap. This behavior is consistent with residual box-geometry uncertainty even when interaction timing is largely preserved.

Table~\ref{tab:multi_case_audit_summary_cr} shows that the audited analysis is not limited to one pair type. Within the same deployment, the reviewed set also contains truck--truck, car--bicycle, car--car, and truck--car interactions. These cases are still too few for a full statistical study, but they help demonstrate that the audit pipeline is being exercised across multiple interaction types rather than only the anchor truck--bicycle example.

\label{sec:longitudinal_ttc_cr}
To interpret the interaction dynamics beyond direction-agnostic safety screening, we compute the longitudinal time-to-collision (TTC$_{\parallel}$) for the heavy vehicle. Let $\mathbf{v}_h$ denote the heavy vehicle velocity and $\hat{\mathbf{e}}_{\parallel}=\mathbf{v}_h/\|\mathbf{v}_h\|$ its unit heading direction. For bicycle position $\mathbf{p}_b$ and heavy-vehicle position $\mathbf{p}_h$,
\[
d_{\parallel}(t) = (\mathbf{p}_b(t)-\mathbf{p}_h(t)) \cdot \hat{\mathbf{e}}_{\parallel},
\]
\[
v_{\parallel}(t) = (\mathbf{v}_b(t)-\mathbf{v}_h(t)) \cdot \hat{\mathbf{e}}_{\parallel},
\]
\[
\mathrm{TTC}_{\parallel}(t) =
\begin{cases}
\dfrac{d_{\parallel}(t)}{-v_{\parallel}(t)}, & v_{\parallel}(t) < 0, \\
+\infty, & \text{otherwise}.
\end{cases}
\]

\begin{figure}[t]
\centering
\includegraphics[width=0.82\columnwidth]{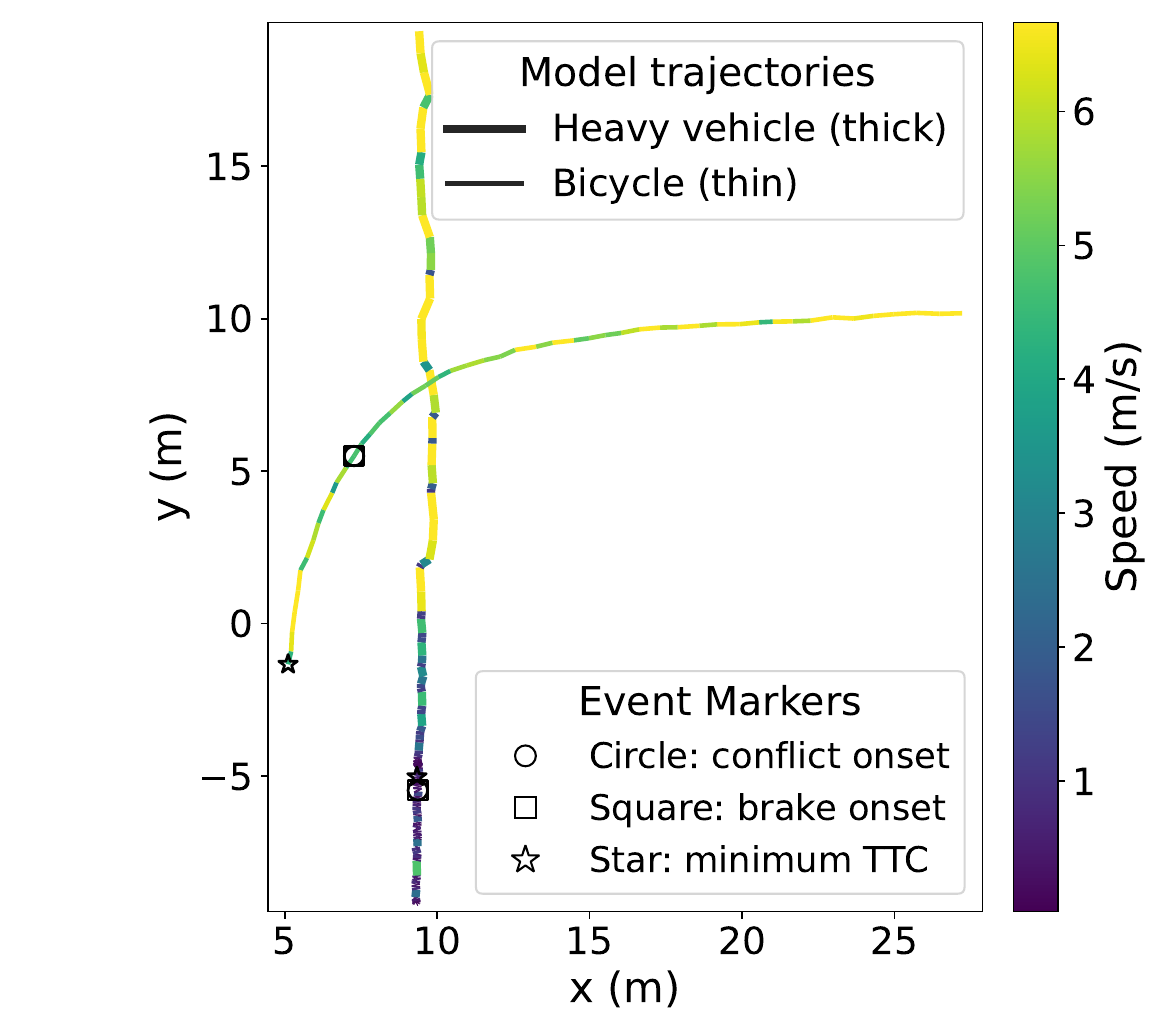}
\caption{Spatiotemporal interaction window for the anchor heavy vehicle--bicycle encounter. Trajectories are colored by speed, and event markers indicate conflict onset, brake onset, and minimum TTC.}
\label{fig:spatiotemporal_window_cr}
\end{figure}

\begin{figure}[t]
\centering
\includegraphics[width=\linewidth]{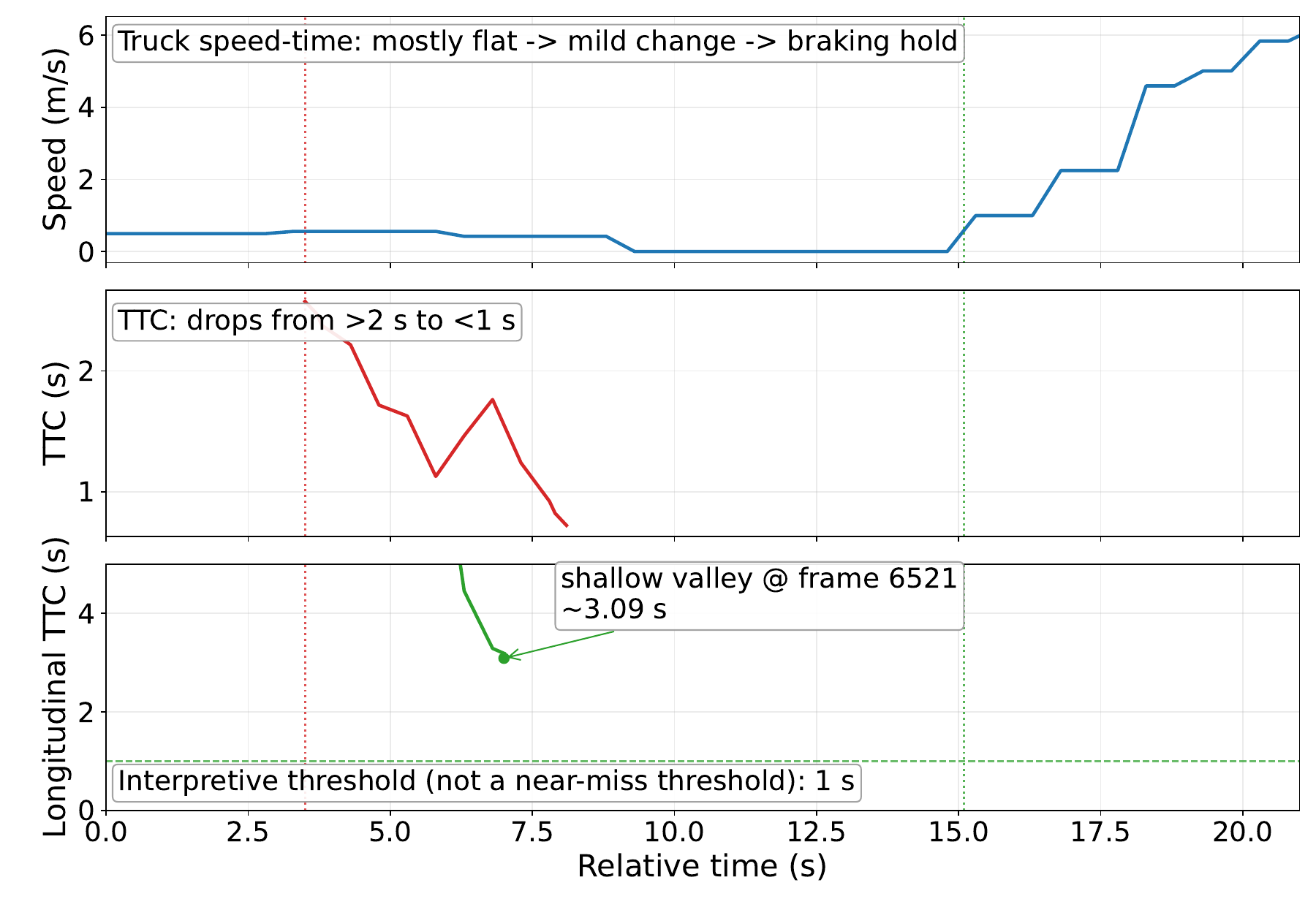}
\caption{Anchor heavy vehicle--bicycle interaction. The x-axis shows relative time in seconds from the start of the audited interaction window. Top: heavy-vehicle speed. Middle: direction-agnostic TTC rapidly drops below 1\,s. Bottom: longitudinal TTC remains above conservative braking thresholds, suggesting a lateral-intrusion-dominated interaction.}
\label{fig:curve_shapes_cr}
\end{figure}

Figure~\ref{fig:spatiotemporal_window_cr} shows the interaction geometry in space and time, while Fig.~\ref{fig:curve_shapes_cr} summarizes the corresponding surrogate-safety curves. Direction-agnostic TTC decreases sharply from above 2\,s to a minimum near 0.62\,s, while longitudinal TTC shows only a shallow minimum around 3.1\,s. This discrepancy suggests that the near-miss is not primarily constrained by the heavy vehicle's longitudinal braking authority. Instead, it is more consistent with rapid lateral intrusion by the bicycle, which compresses the available safety margin despite sufficient stopping capacity along the truck heading. We therefore use longitudinal TTC here as a mechanism-explanation tool, not as a replacement for the primary event-mining criterion.

At larger scale within the same deployment, automatically mined movement summaries over the first 16,000 predicted frames yielded 336 movement-valid tracks, 6,046 quality-filtered candidate pairs, and 604 near-miss events. We use this aggregate only as a recurrence indicator rather than a human-validated benchmark, but it supports the view that the anchor case is not an isolated artifact of one short window.

\begin{figure}[t]
\centering
\includegraphics[width=0.85\linewidth]{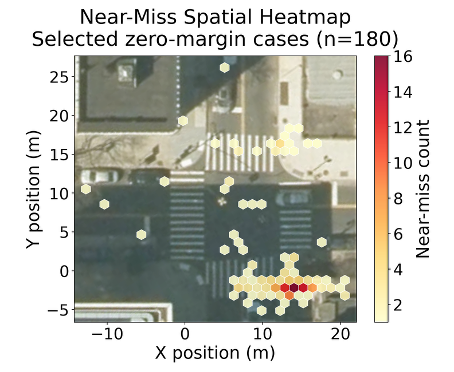}
\caption{Post–8,000-frame vehicle–VRU near-miss overlay mined from subsequent roadside LiDAR data outside the labeled set. The label $n=180$ denotes the number of selected zero-margin cases plotted in this heatmap. Automatically detected events remain concentrated within the same central conflict zone.}
\label{fig:post8k_near_miss_overlay_cr}
\end{figure}

Figure~\ref{fig:post8k_near_miss_overlay_cr} provides the qualitative recurrence view behind that aggregate summary. Although these later events are not human-validated one by one, the hotspot concentration persists in the same central intersection region rather than dispersing uniformly across the scene. That recurrence supports the broader claim that roadside sensing can expose site-specific conflict structure over time, even when the present paper remains focused on post-hoc auditing rather than real-time intervention.

\section{Discussion and Conclusion}
This pilot shows how roadside LiDAR can be used as an external safety-audit reference at a complex urban intersection. The main contribution is not a new perception backbone, but a workflow that makes infrastructure-side trajectories more reviewable through conservative refinement, repeated QA, and mechanism-aware near-miss interpretation.

The evaluation remains intentionally limited. Although we include detector results against human-reviewed ground truth, audited branch comparisons, stitched-window tracking evaluation, and human cross-checks, the paper does not yet provide a large-scale event-level benchmark across vehicle classes, traffic conditions, or multiple sites. The present claims should therefore be read as evidence of feasibility, auditability, and initial explanatory value rather than definitive reliability or full-stack trustworthiness guarantees.

Future work should expand the current pipeline to larger multi-intersection studies with more extensive model-versus-human near-miss validation, stronger identity-consistent tracking metrics, and explicit treatment of synchronization, communication, and calibration constraints. A longer journal version is ongoing work and will elaborate the broader dataset comparison, full QA dashboard evidence, post-8,000 hotspot recurrence analysis, and additional engineering extensions such as learned residual stabilization.

{\small
\bibliographystyle{ieeenat_fullname}
\bibliography{references}

@article{ji2024cvis_survey,
  title        = {Toward autonomous vehicles: A survey on cooperative vehicle-infrastructure system},
  author       = {Ji, Y. and others},
  journal      = {iScience},
  year         = {2024},
  url          = {https://www.sciencedirect.com/science/article/pii/S2589004224009738},
  doi          = {10.1016/j.isci.2024.109751}
}

@misc{huang2023v2xcp_survey,
  title        = {Vehicle-to-Everything Cooperative Perception for Autonomous Driving: A Survey},
  author       = {Huang, T. and others},
  year         = {2023},
  howpublished = {arXiv preprint},
  url          = {https://arxiv.org/abs/2310.03525}
}

@inproceedings{yu2022dairv2x,
  title     = {DAIR-V2X: A Large-Scale Dataset for Vehicle-Infrastructure Cooperative 3D Object Detection},
  author    = {Yu, Haibao and others},
  booktitle = {Proceedings of the IEEE/CVF Conference on Computer Vision and Pattern Recognition (CVPR)},
  year      = {2022},
  url       = {https://openaccess.thecvf.com/content/CVPR2022/html/Yu_DAIR-V2X_A_Large-Scale_Dataset_for_Vehicle-Infrastructure_Cooperative_3D_Object_Detection_CVPR_2022_paper.html}
}

@inproceedings{yu2023v2xseq,
  title     = {V2X-Seq: A Large-Scale Sequential Dataset for Vehicle-Infrastructure Cooperative Perception and Forecasting},
  author    = {Yu, Haibao and others},
  booktitle = {Proceedings of the IEEE/CVF Conference on Computer Vision and Pattern Recognition (CVPR)},
  year      = {2023},
  url       = {https://openaccess.thecvf.com/content/CVPR2023/html/Yu_V2X-Seq_A_Large-Scale_Sequential_Dataset_for_Vehicle-Infrastructure_Cooperative_Perception_and_CVPR_2023_paper.html}
}

@article{roadside_lidar_review2026,
  title   = {Roadside lidar-based scene understanding toward intelligent traffic perception: A comprehensive review},
  author  = {Zhang, Jiaxing and Ge, Chengjun and Xiao, Wen and Tang, Miao and Mills, Jon and Coifman, Benjamin and Chen, Nengcheng},
  journal = {ISPRS Journal of Photogrammetry and Remote Sensing},
  year    = {2026},
  doi     = {10.1016/j.isprsjprs.2026.01.012},
  url     = {https://www.sciencedirect.com/science/article/pii/S0924271626000122}
}

@article{lin2023roadside_tracking,
  title   = {Vehicle Detection and Tracking with Roadside LiDAR Using Low-Latency Edge Computing},
  author  = {Lin, C. and others},
  journal = {Sensors},
  year    = {2023},
  url     = {https://pmc.ncbi.nlm.nih.gov/articles/PMC10575351/},
  doi     = {10.3390/s23198125}
}

@article{mo2024enhanced_perception_intersections,
  title   = {Enhanced Perception for Autonomous Vehicles at Intersections Using Roadside Sensing and V2X Communication},
  author  = {Mo, Y. and others},
  journal = {Sensors},
  year    = {2024},
  url     = {https://pmc.ncbi.nlm.nih.gov/articles/PMC10856888/},
  doi     = {10.3390/s24030931}
}

@techreport{gettman2003ssm,
  title       = {Surrogate Safety Measures From Traffic Simulation Models},
  author      = {Gettman, D. and Head, L.},
  institution = {Federal Highway Administration (FHWA)},
  year        = {2003},
  number      = {FHWA-RD-03-050},
  url         = {https://ntlrepository.blob.core.windows.net/lib/38000/38000/38015/FHWA-RD-03-050.pdf},
  note        = {HTML landing page: https://www.fhwa.dot.gov/publications/research/safety/03050/07.cfm}
}

@misc{hyden1987tct,
  title        = {The Swedish Traffic Conflicts Technique},
  author       = {Hyd{\'e}n, Christer},
  year         = {1987},
  howpublished = {Technical report / calibration study},
  url          = {https://www.ictct.net/wp-content/uploads/SMoS_Library/LIB_Hyden_1987.pdf}
}

@article{johnsson2021relative_validation,
  title   = {A relative approach to the validation of surrogate measures of safety},
  author  = {Johnsson, C. and others},
  journal = {Accident Analysis \& Prevention},
  year    = {2021},
  url     = {https://www.sciencedirect.com/science/article/pii/S000145752100381X},
  doi     = {10.1016/j.aap.2021.106511}
}

@inproceedings{zimmer2024tumtraf_v2x,
  title     = {TUMTraf V2X Cooperative Perception Dataset},
  author    = {Zimmer, Walter and others},
  booktitle = {Proceedings of the IEEE/CVF Conference on Computer Vision and Pattern Recognition (CVPR)},
  year      = {2024},
  url       = {https://openaccess.thecvf.com/content/CVPR2024/papers/Zimmer_TUMTraf_V2X_Cooperative_Perception_Dataset_CVPR_2024_paper.pdf},
  note      = {Project page: https://tum-traffic-dataset.github.io/tumtraf-v2x/ ; arXiv: https://arxiv.org/abs/2403.01316}
}

@article{yazgan2024cp_datasets_survey,
  title        = {Collaborative Perception Datasets in Autonomous Driving},
  author       = {Yazgan, M. and others},
  year         = {2024},
  howpublished = {arXiv preprint},
  url          = {https://arxiv.org/abs/2404.14022}
}

@misc{urbaning2025urbaning_v2x,
  title        = {UrbanIng-V2X: A Large-Scale Multi-Vehicle, Multi-Infrastructure Dataset Across Multiple Intersections for Cooperative Perception},
  author       = {Anonymous},
  year         = {2025},
  howpublished = {arXiv preprint},
  url          = {https://arxiv.org/abs/2510.23478}
}

@phdthesis{jansson2005collision_avoidance,
  title   = {Collision Avoidance Theory: with Application to Automotive Collision Mitigation},
  author  = {Jansson, Jonas},
  school  = {Link{\"o}ping University},
  address = {Link{\"o}ping, Sweden},
  year    = {2005},
  type    = {PhD dissertation},
  url     = {https://urn.kb.se/resolve?urn=urn:nbn:se:liu:diva-91385}
}

@article{ward2015extending_ttc,
  title   = {Extending Time to Collision for probabilistic reasoning in general traffic scenarios},
  author  = {Ward, James R. and Agamennoni, Gabriel and Worrall, Stewart and Bender, Asher and Nebot, Eduardo},
  journal = {Transportation Research Part C: Emerging Technologies},
  volume  = {51},
  pages   = {66--82},
  year    = {2015},
  doi     = {10.1016/j.trc.2014.11.002}
}

@inproceedings{yin2021centerpoint,
  title     = {Center-Based 3D Object Detection and Tracking},
  author    = {Yin, Tianwei and Zhou, Xingyi and Kr{\"a}henb{\"u}hl, Philipp},
  booktitle = {Proceedings of the IEEE/CVF Conference on Computer Vision and Pattern Recognition (CVPR)},
  pages     = {11784--11793},
  year      = {2021},
  doi       = {10.1109/CVPR46437.2021.01161}
}

@inproceedings{bewley2016sort,
  title     = {Simple Online and Realtime Tracking},
  author    = {Bewley, Alex and Ge, Zongyuan and Ott, Lionel and Ramos, Fabio and Upcroft, Ben},
  booktitle = {2016 IEEE International Conference on Image Processing (ICIP)},
  pages     = {3464--3468},
  year      = {2016},
  doi       = {10.1109/ICIP.2016.7533003}
}
}
\end{document}